\documentclass[
 reprint,
 amsmath,
 amssymb,
 aps,
pra,
]{revtex4-2}

\usepackage{textcomp}
\usepackage{wrapfig}
\usepackage{float}
\usepackage{amsmath}
\usepackage{cleveref}
\usepackage[dvipsnames]{xcolor}
\usepackage{graphicx}
\usepackage{dcolumn}
\usepackage{bm}
\usepackage{booktabs}

\usepackage{physics}

\begin{document}

\title{Numerical simulation methods for quantum sensing {at parametric criticality}}

\author{Kirill Petrovnin}
\email{kirill.petrovnin@aalto.fi}
\affiliation{QTF  Centre  of  Excellence, 
	Department of Applied Physics, Aalto University, FI-00076 Aalto, Finland\\}
\author{Jiaming Wang}
\affiliation{QTF  Centre  of  Excellence, 
	Department of Applied Physics, Aalto University, FI-00076 Aalto, Finland\\}
\author{Gheorghe Sorin Paraoanu}
\affiliation{QTF  Centre  of  Excellence, 
	Department of Applied Physics, Aalto University, FI-00076 Aalto, Finland\\}

             \begin{abstract}
{Microwave photon detection is a key technology for low-temperature superconducting electronics and quantum information processing. A promising possibility is to use switching processes in parametric superconducting devices at criticality, which can be triggered by small perturbations.  Here we} demonstrate the unique sensing properties of the {superconducting} Kerr parametric resonator {when operated in the proximity of the phase transition boundary}. We utilize a semiclassical approximation to provide numerical and analytical results for the Heisenberg-Langevin and Fokker-Planck equations that describe the switching mechanism. We show that the probability of switching events is enhanced by probe input states with {energies down to single quanta levels.}
\end{abstract}
             
\maketitle

\section*{Introduction}

Parametric phenomena typically {emerge in situations when the parameter that determine the restoring force acting on a harmonic oscillator is modulated} \cite{eichlerClassicalQuantumParametric2023a}. In mechanical systems, for instance, a child on a swing can sustain motion by alternating between standing and squatting, thereby modulating the effective length of the pendulum. In electrical systems, parametric resonance is achieved by periodically varying the reactance (capacitance or inductance) of a resonant circuit, a principle widely applied in low-noise amplification of weak signals. Notably, {in low-temperature nanoelectronics this has led to the realization of} superconducting parametric devices such as Josephson parametric amplifiers (JPAs) \cite{castellanos-beltranWidelyTunableParametric2007, yamamotoFluxdrivenJosephsonParametric2008} and Josephson traveling wave parametric amplifiers (JTWPAs) \cite{macklinQuantumLimitedJosephson2015}, which are extensively used for superconducting qubit readout \cite{aumentadoSuperconductingParametricAmplifiers2020} and the generation of entangled states \cite{eichlerObservationTwoModeSqueezing2011, bergealTwoModeCorrelationMicrowave2012, menzelPathEntanglementContinuousVariable2012, petrovninGenerationStructuringMultipartite2023}. The periodic signal used to modulate the parameters of the system and supply the energy for amplification is typically called the pump. 
{As the pump strength increases, it reaches a critical threshold where the energy input compensates for losses \cite{wustmann2019parametric, eichlerClassicalQuantumParametric2023a}. Beyond this point, the system enters a regime of parametric instability, characterized by self-sustained oscillations at the resonance frequency, even in the absence of an external input.} 
A parametric device {thus} functions as an amplifier only when the pump strength remains below this critical threshold. {The onset of these oscillations is commonly interpreted as a non-equilibrium phase transition in a driven-dissipative system---a phenomenon that can be regarded as a finite component phase transition \cite{wilsonPhotonGenerationElectromagnetic2010, wustmannParametricResonanceTunable2013}. } 
These phase transitions can also be harnessed for detection applications, {because near the criticality the system becomes very sensitive to small perturbations at single-quanta levels} \cite{dicandiaCriticalParametricQuantum2023, alushiOptimalityNoiseResilience2024, chavez-carlosQuantumSensingKerr2024}.

{The study of dynamics in parametric driven-dissipative Kerr systems is also of fundamental importance, as they offer a unifying framework for numerous quantum and condensed-matter models. 
For example, critical slowing down and its associated timescale can be studied experimentally \cite{Scarlino_2025}. Such systems can also be utilized as parametric phase-locked oscillators (PPLOs) in the microwave range, which can be regarded as a fundamental logic element 
\cite{Yamamoto_2014}. Even in the absence of parametric pumping, the nonlinearity is also interesting, as the physics of the quantum Duffing oscillator emerges if the system is linearly driven \cite{Gross_2023}.
In addition, two recent works in optics highlight the potential of operating the parametric oscillator as a controllable random bit generator that enables photonic probabilistic computing and machine learning \cite{roques-carmesBiasingQuantumVacuum2023, choiPhotonicProbabilisticMachine2024}.}

In this work, we model our system as a nonlinear parametric (Kerr) oscillator, where the time evolution is governed by the Heisenberg-Langevin equations for the in-phase and quadrature operators.
{In the regime where the nonlinearity is much smaller than the bandwidth, a semiclassical approach is possible since photon blockade effects are negligible \cite{PhysRevApplied.13.044017,linCriticalFluctuationsRates2015}}.
By numerically solving these equations of motion, we evaluate key performance indicators, such as detection efficiency and dark count probability, when operating our device as a photon detector.
Moreover, near the critical threshold, the  dynamics can be described {semiclassically} by the evolution of a slow variable, or ``soft mode'' \cite{dykmanFluctuatingNonlinearOscillators2012}, through introducing an effective potential \cite{zorinPerioddoublingBifurcationReadout2011, dykmanFluctuatingNonlinearOscillators2012, linCriticalFluctuationsRates2015}, which is analogous to Landau's free energy in phase transition theory \cite{linCriticalFluctuationsRates2015}. 

The results presented here pave the way for applications in single-photon detection and integration into more complex quantum systems, such as those involving superconducting and spin qubits.

\section*{Results and Discussion}

\subsection*{Theoretical model}
We start with an overview of the dynamics of the parametrically pumped harmonic oscillator with Kerr nonlinearity driven by a probe coherent field and by noise.
Since our primary interest lies in obtaining the time evolution of the expectation value of the cavity mode $a$ (or equivalently, quadratures $\mathcal{Q}$ and $\mathcal{P}$), rather than the density matrix, we formulate the problem in the Heisenberg picture and solve the Heisenberg-Langevin equations of motion by using a semiclassical approximation.  Our treatment follows the analysis based on the separation of slow and fast variables  \cite{dykmanFluctuatingNonlinearOscillators2012,linCriticalFluctuationsRates2015} supplemented with the input-output theory for the driving field \cite{petrovnin_microwave_2024}.
It should be noted that this approach does not include the up/down-converted quantum noise resulting from mode mixing \cite{PhysRevApplied.13.044017}. {This allows us to describe the dynamics as a diffusive motion in an effective potential.} \\

\subsubsection*{Heisenberg-Langevin equations of motion}

In the most general form, the system Hamiltonian is 
\begin{equation}
	H_{\rm sys} = \hbar\omega_{\rm 0} a^{\dag}a + \frac{\hbar}{2}\left[ \alpha e^{i \omega_{\rm P}t} + c.c.\right](a+a^{\dag})^2 + \hbar K (a + a^\dag)^4 , \label{eq:initial}
\end{equation}
where $\omega_{0}$ is the frequency of the harmonic oscillator, $\alpha =|\alpha|\exp(i \theta_{\rm P})$ is the complex parametric coupling and $K <0$ is the Kerr constant. 
In terms of the standard harmonic oscillator position $q$ and momentum $p$, where
\begin{equation}
	q=\sqrt{\frac{\hbar}{2\omega_{\rm 0}}}(a + a^{\dag}) 
	~~~{\rm and}~~~
	p=\frac{1}{i}\sqrt{\frac{\hbar \omega_{\rm 0}}{2}}(a - a^{\dag}) ,
\end{equation} 
and up to a constant $\hbar \omega_{\rm 0}/2$, this Hamiltonian reads 
\begin{equation}
	H_{\rm sys} = \frac{p^2}{2} + \frac{1}{2} \omega_{\rm 0}^2 q^2 + 
	2 |\alpha| \omega_{\rm 0} q^2 \cos(\omega_{\rm P}t + \theta_{\rm P}) + \frac{4K\omega_{\rm 0}^2}{\hbar} q^4 .\nonumber
\end{equation}
However, it is more convenient to use the dimensionless quadratures
\begin{equation}
\label{eq:Q_P_notation}
	\mathcal{Q}=\frac{1}{\sqrt{2}}(a + a^{\dag})~~~{\rm and}~~~
	\mathcal{P}=\frac{1}{i\sqrt{2}}(a - a^{\dag}),
\end{equation}    
satisfying $(\mathcal{P}^2 + \mathcal{Q}^2)/2 = a^{\dag}a + 1/2$ and have the system Hamiltonian written up in the form
\begin{equation}
	H_{\rm sys} = \frac{\hbar \omega_{\rm 0}}{2} \left(\mathcal{P}^2 + \mathcal{Q}^2\right) + 2 \hbar |\alpha| \mathcal{Q}^2 \cos(\omega_{\rm P}t + \theta_{\rm P}) + 4\hbar K \mathcal{Q}^4.
\end{equation}
Next, we move in a rotating frame defined by $U = \exp [-i(\omega_{\rm P}t/2 + \theta /2)a^{\dag}a]$, where $\theta$ is for the moment an arbitrary phase, which results in the transformed quadratures
\begin{align}
	U^{\dagger}\mathcal{Q}U&=\mathcal{Q}\cos\left(\frac{\omega_{\rm P}t}{2}+ \frac{\theta}{2}\right) +
	\mathcal{P}\sin\left(\frac{\omega_{\rm P}t}{2}+ \frac{\theta}{2}\right) ,\\
	U^{\dagger}\mathcal{P}U&=-\mathcal{Q}\sin\left( \frac{\omega_{\rm P}t}{2}+ \frac{\theta}{2}\right) +
	\mathcal{P}\cos\left(\frac{\omega_{\rm P}t}{2}+ \frac{\theta}{2}\right) .
\end{align}
Here we have used the well-known identity $e^{i \xi a^{\dag}a} a e^{-i \xi a^{\dag}a} = e^{-i \xi}a$, which yields
$U^{\dag}a U = e^{-\frac{i}{2} \left(\omega_{\rm P}t + \theta \right)} a$ and $U^{\dag}a^{\dag} U = e^{\frac{i}{2} \left(\omega_{\rm P}t + \theta \right)} a^{\dag}$.

In consequence, the transformed Hamiltonian is $U^{\dagger}H_{\rm sys}U -i \hbar U^{\dagger}\dot{U}$, and, by eliminating the fast rotating terms we get the rotating-wave approximation (RWA) system Hamiltonian  
\begin{align}
	H_{\rm sys}^{\rm (RWA)} &=& \frac{\hbar \Delta}{2}(\mathcal{P}^2 + \mathcal{Q}^2)
	+ \frac{\hbar |\alpha |}{2}[(-\mathcal{P}^2 + \mathcal{Q}^2)\cos (\theta -\theta_{\rm P})   \nonumber \\
	& &  + (\mathcal{P}\mathcal{Q} +
	\mathcal{Q}\mathcal{P} )\sin (\theta -\theta_{\rm P})] + \frac{3\hbar K}{2} (\mathcal{P}^2 + \mathcal{Q}^2)^2 ,
	\label{H_QP}
\end{align}
where we have introduced the detuning of the oscillator frequency with respect to half the pump frequency as $\Delta\equiv\omega_0-\omega_{\rm P}/2$.
Alternatively, by using the annihilation and creation operators, we can write
\begin{align}
	{H}_{\text{sys}}^{\rm (RWA)} =& {\hbar}(\Delta + 12K)  a^\dagger a + \frac{\hbar |\alpha|}{2}\left[a^{2}e^{-i(\theta -\theta_{\rm P})}  + a^{\dagger 2}e^{i(\theta -\theta_{\rm P})}\right] \nonumber \\ &
    + 6{\hbar}{K} a^{\dagger} a^{\dagger} a a  .\label{eq:HRWA}
\end{align}

Next, we study what happens when the harmonic oscillator becomes an open system. Formally, this is done through introducing a coupling rate $\gamma$ to an internal environment and a coupling rate $\kappa$ to an external environment. The internal environment is assumed to be a thermal bath, while the external environment contains a thermal bath and also a drive field, assumed to be coherent. The latter can be modeled as a classical field $b(t) = b\exp \left (-i \omega t\right) = |b|\exp \left (-i \omega t - i \varphi\right)$ that couples into the oscillator via the external port, resulting in the probe Hamiltonian 
\begin{align}
	H_{p} & =  -2\sqrt{2}\hbar \sqrt{\kappa }|b| \cos \left(\omega t + \varphi \right) \mathcal{P} \\ & = 	i \hbar \sqrt{\kappa}  \left[b(t)^{*} + b(t)\right]\left(a - a^{\dag}\right),
\end{align}
The average number of photons in this field in a time interval $\tau$ is 
$\bar{n} = |b|^2 \tau$. We take the frequency of this field as $\omega = \omega_{\rm P}/2$ to get in the rotating frame defined by $U$:
\begin{equation}
	H_{p}^{\rm (RWA)} = i \hbar \sqrt{\kappa} |b| \left[e^{-i \left(\frac{\theta}{2}-\varphi \right)}a - e^{i \left(\frac{\theta}{2} - \varphi \right)}a^{\dag} 
	\right] ,
\end{equation}
the result of applying the RWA to the rotated Hamiltonian $U^{\dag}H_{p}U$.
Alternatively, by employing $\mathcal{Q}$ and $\mathcal{P}$, we can write
\begin{equation}
	H_{p}^{\rm (RWA)} =\sqrt{2\kappa}|b|\hbar \left[\mathcal{Q} \sin \left(\frac{\theta}{2} -\varphi \right) - \mathcal{P} \cos \left(\frac{\theta}{2}-\varphi \right)\right].
\end{equation}
We can then construct a Hamiltonian describing a coherently driven Kerr harmonic oscillator with parametric pumping  
\begin{equation}
	H_{b}^{\text{(RWA)}} = H_{\rm sys}^{\text{(RWA)}}  +  H_{p}^{\text{(RWA)}}.
    \label{eq:HRWA_Hp}
\end{equation}
For the thermal noise we assume the same temperature $T$ and average number of particles per mode $\bar{n}_{T}$ for both the external and internal modes.  
{Here $\bar{n}_T= (e^\frac{\hbar \omega_{}}{k_{\rm B} T}-1)^{-1}$ is the mode occupation number at the resonator frequency.} The noise operators $\xi_{\mathcal{Q}_{\rm in}}$ and $\xi_{\mathcal{P}_{\rm in}}$ have the correlations
\begin{eqnarray}
	\langle  \xi_{\mathcal{Q}_{\rm in}}(t)  \xi_{\mathcal{Q}_{\rm in}}(t')  \rangle &=& (\bar{n}_{T}+ 1/2) \delta (t-t'), \label{eq:noiseqq} \\
	\langle  \xi_{\mathcal{P}_{\rm in}}(t)  \xi_{\mathcal{P}_{\rm in}}(t')  \rangle &=& 
	(\bar{n}_{T}+ 1/2) \delta (t-t') , \label{eq:noisepp} \\
		\langle  \left[\xi_{\mathcal{Q}_{\rm in}}(t),  \xi_{\mathcal{P}_{\rm in}}(t') \right] \rangle &=& i \delta (t - t') . \label{eq:noisepq}
\end{eqnarray}
With these notations we can write the Heisenberg-Langevin equations
\begin{align}
  \dot{\mathcal{Q}} &= \frac{i}{\hbar}[H_{b}^{\text{(RWA)}},\mathcal{Q}]  - \frac{\kappa + \gamma}{2} \mathcal{Q}
  - \sqrt{\kappa + \gamma} \xi_{\mathcal{Q}_{\rm in}}, \label{eq:HL1}\\
    \dot{\mathcal{P}} &= \frac{i}{\hbar}[H_{b}^{\text{(RWA)}},\mathcal{P}]  - \frac{\kappa + \gamma}{2} \mathcal{P}
  - \sqrt{\kappa + \gamma} \xi_{\mathcal{P}_{\rm in}},\label{eq:HL2}
\end{align}
or explicitly
\begin{eqnarray}
	\dot{\mathcal{Q}} &=& \left[|\alpha |\sin(\theta - \theta_{\rm P}) - \frac{\kappa + \gamma}{2} \right] \mathcal{Q} + \left[\Delta -|\alpha |\cos(\theta - \theta_{\rm P})\right] \mathcal{P}  \nonumber \\
	& &+ 6 K \left[ \frac{1}{2}(\mathcal{P}\mathcal{Q}^2 + \mathcal{Q}^2\mathcal{P}) + \mathcal{P}^3\right] - \sqrt{2\kappa} |b|\cos\left(\frac{\theta}{2}-\varphi \right) \nonumber \\
	& & - \sqrt{\kappa + \gamma} \xi_{\mathcal{Q}_{\rm in}} 
	\label{eq:probedotq},\\
	\dot{\mathcal{P}} &=& -\left[|\alpha |\sin(\theta - \theta_{\rm P}) + \frac{\kappa + \gamma}{2} \right] \mathcal{P} - \left[\Delta +|\alpha |\cos(\theta - \theta_{\rm P})\right] \mathcal{Q}  \nonumber \\
	& &- 6 K \left[ \frac{1}{2}(\mathcal{Q}\mathcal{P}^2 + \mathcal{P}^2\mathcal{Q}) + \mathcal{Q}^3\right] - \sqrt{2\kappa} |b|\sin\left(\frac{\theta}{2}-\varphi \right) \nonumber \\
	& & - \sqrt{\kappa + \gamma} \xi_{\mathcal{P}_{\rm in}}. \label{eq:probedotp}
\end{eqnarray}
{Note that in the equation above we retain the order of operators in products such as $\mathcal{P}\mathcal{Q}^2$, $\mathcal{Q}^2\mathcal{P}$ since $\mathcal{P}$ and $\mathcal{Q}$ do not commute, $[\mathcal{Q}, \mathcal{P}] = i$. Also, we can regard the term proportional with the $b$ field as resulting from the action of a displacement operator on the external input field \cite{RevModPhys.82.1155}. This separates the coherent contribution from the remaining fluctuating fields $\xi_{\mathcal{Q}_{\rm in}}$, $\xi_{\mathcal{P}_{\rm in}}$ with correlations given by
Eqs. (\ref{eq:noiseqq},\ref{eq:noisepp},\ref{eq:noisepq}).
}

\subsubsection*{Separation of slow and fast quadratures and the effective potential theory}
\label{sec:IIB}

{We solve the Heisenberg-Langevin equations in the semiclassical approximation, where $\mathcal{P}$ and $\mathcal{Q}$ are replaced by classical dynamical variables. This approximation can be justified by several arguments. For example, in \cite{linCriticalFluctuationsRates2015} an effective Planck constant is obtained, and a semiclassical approach is deemed to be valid when this quantity becomes much smaller than 1. 
For our system, this translates into $K \ll (\kappa + \gamma)$, which is well satisfied. In this situation photon-blockade effect are negligible -- the system is far from ``qubit-like'' behaviour \cite{PhysRevApplied.13.044017}.
Another argument can be obtained by examining Eqs. (\ref{eq:probedotq}, \ref{eq:probedotp}) while replacing $\mathcal{P}$ and $\mathcal{Q}$ by their averages on a coherent state. Then the needed re-orderings 
will affect only the terms containing $\mathcal{P}^2\mathcal{Q}$, $\mathcal{Q}^2\mathcal{P}$ etc. generating additional terms that are first order in $\mathcal{P}$ and $\mathcal{Q}$, which will renormalize the first two right-hand-side terms of the Heisenberg-Langevin equation. However, the effect of these terms is negligible if $K$ is the smallest parameter in Eqs. (\ref{eq:probedotq}, \ref{eq:probedotp}), that is $K\ll (\kappa + \gamma), |\alpha |, \Delta$, etc., which is indeed the case for our system. }

To proceed, we employ the method of separation of slow and fast variables \cite{linCriticalFluctuationsRates2015}, noticing that if $\theta - \theta_{\rm P} \approx \pi/2$, then $\mathcal{P}$ becomes a fast variable that follows the slower variable $\mathcal{Q}$.
This can be seen from the linearized version of Eqs. (\ref{eq:probedotq},\,\ref{eq:probedotp}), in which, recalling that we operate sufficiently close to criticality,  $\mathcal{P}$ rotates with angular frequency $|\alpha|+ (\kappa + \gamma)/2$, while $\mathcal{Q}$ rotates at a much slower frequency $|\alpha|- (\kappa + \gamma)/2$. 
Therefore, we can put $\dot{\mathcal{P}}=0$ and Eq. (\ref{eq:probedotp}) reads:
\begin{align}
	\mathcal{P} \approx &- \frac{1}{|\alpha | \sin (\theta - \theta_{\rm P} )  + \frac{\kappa + \gamma}{2}}\big[(\Delta + |\alpha| \cos (\theta - \theta_{\rm P}))\mathcal{Q} + \nonumber\\ 
    & + 6 K\mathcal{Q}^3\big] - \frac{\sqrt{2\kappa} |b| \sin\left(\frac{\theta}{2} - \varphi\right)}{|\alpha| \sin(\theta - \theta_{\rm P}) + \frac{\kappa + \gamma}{2}}. \nonumber
	\end{align}	
{Note that this approximation manifestly breaks the commutation relations between the quadratures, therefore it is semiclassical.}
We insert this expression into Eq. (\ref{eq:probedotq}) and neglect small nonlinear terms. 
When doing so, the coherent drive produces a term $\propto \sin (\theta /2-\varphi)$ and another one $\propto \cos (\theta /2-\varphi)$. We keep only the latter since it is larger than the first one by a factor $(\Delta - |\alpha|\cos (\theta - \theta_{\rm P}))/(|\alpha | \sin (\theta - \theta_{\rm P})   + \frac{\kappa + \gamma}{2}) \ll 1$. As a final result we  obtain
\begin{equation}
	\dot{\mathcal{Q}} =	- \partial_{\mathcal{Q}}\mathcal{U}_{b}(\mathcal{Q}) - \sqrt{\kappa + \gamma} \xi_{\mathcal{Q}_{\rm in}} .
\end{equation}
where
	\begin{equation}
		\mathcal{U}_{b}(\mathcal{Q}) =  \mathcal{U} (\mathcal{Q}) + \sqrt{2\kappa }|b| \cos \left(\frac{\theta}{2} - \varphi \right) \mathcal{Q}.
	\end{equation}
Here the effective potential $\mathcal{U}_{b=0}(\mathcal{Q}) =\mathcal{U}(\mathcal{Q})$ in the absence of coherent driving is 
\begin{equation}
	\mathcal{U}(\mathcal{Q})  =  \frac{2}{|\alpha|\sin(\theta - \theta_{\rm P}) + \frac{\kappa + \gamma}{2}} \times \mathcal{V} (\mathcal{Q})  ,\label{eq:eff} 
\end{equation}	
with
\begin{equation}	
	\mathcal{V} (\mathcal{Q}) =	\left[ \left(\frac{\kappa + \gamma}{2}\right)^2 + \Delta^2 - |\alpha|^2 \right]\frac{\mathcal{Q}^2}{4} + 3\Delta K \frac{\mathcal{Q}^4}{2} + 3 K^2 \mathcal{Q}^6 .\nonumber
\end{equation}

Let us define the critical {pump amplitude}
\begin{equation}
\label{eq:alpha_c}
\alpha_{c}(\Delta )= \sqrt{\left(\frac{\kappa + \gamma}{2}\right)^2 + \Delta^2},
\end{equation}
with $\alpha_{c}(0) = (\kappa + \gamma)/2$.
The $\mathcal{Q}$-dependent part of the effective potential reads:
\begin{equation}
	\mathcal{V} (\mathcal{Q}) =	\left[\alpha_{c}(\Delta )^2 - |\alpha|^2 \right]\frac{\mathcal{Q}^2}{4} + 3\Delta K \frac{\mathcal{Q}^4}{2} + 3 K^2 \mathcal{Q}^6.
    \label{eq:effred}
\end{equation}

\subsubsection*{Fokker-Planck equation}
The Fokker-Planck equation resulting from the Heisenberg-Langevin equation is 
\begin{equation}
	\partial_{t} W_{b} = \partial_{\mathcal{Q}} \left( W_{b} \partial_{\mathcal{Q}}  \mathcal{U}_{b} \right) + D \partial_{\mathcal{Q}}^{2}W_{b},
    \label{eq:FPb}
\end{equation}	
where the diffusion constant is 
\begin{equation}
\label{eq:diffusion_coef}
D = \frac{\kappa + \gamma}{2}\left(\bar{n}_{T} + \frac{1}{2}\right).    
\end{equation}

{In the spirit of the fluctuation-dissipation theorem, this is a relation between the diffusion constant (which determines the fluctuations in the quadratures through the Fokker-Planck equation) and the dissipation or response function (characterized by $\kappa$ and $\gamma$). }

\subsection*{Phase diagram}

{We start by investigating the important case $b=0$}.
The reduced effective potential Eq. (\ref{eq:effred}) can be regarded as an analog of the free energy functional in the Landau theory of phase transitions, with the slow variable
$\mathcal{Q}$ playing the role of the order parameter. Recall also that $K<0$. The quadratic term depends on the difference $\alpha_{c}(\Delta )^2 - |\alpha |^2$, similar to the temperature difference $T-T_{0}$ between temperatures that appears in the standard Landau theory. 
The existence of the quartic and sextic terms is possible only if the Kerr nonlinearity is nonzero: otherwise, the potential would be a parabola for $|\alpha |< \alpha_{c}$ (stable parametric oscillator) and an inverted parabola for $|\alpha | > \alpha_{c}$ (unstable parametric oscillator). The quartic term is positive for $\Delta <0$ and negative for $\Delta >0$. Then, just like in the case of Landau theory, we expect a second-order transition in the first case and a first-order transition in the latter. Finally, the sextic term is always positive, ensuring thermodynamical (dynamical in our case) stability. {One often refers to these transitions as finite-component, in order to distingush them from the regular many-particle versions. Also, to bring up these analogies, we refer to the parametric stable and unstable states as phases and to the parametric instability threshold separating them as phase transition boundary.}

\begin{figure}
    \includegraphics[width=1\linewidth]{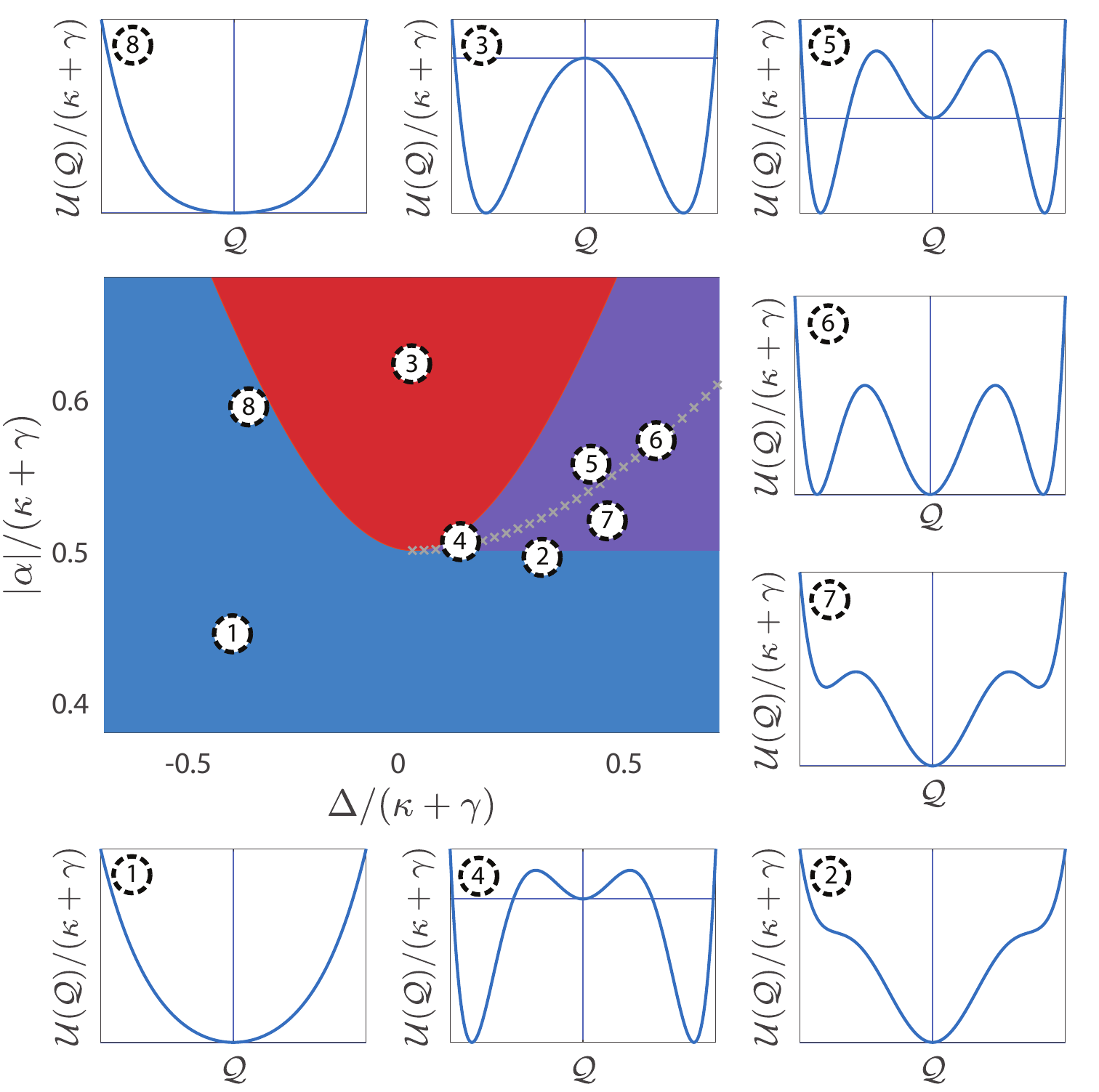}
    \caption{{\bf Phase diagram and the effective potential at various points of detuning $\Delta$ and pump strength $|\alpha|$}.  {The three different colors} {in the phase diagram} {correspond to different stability regions of the system, defined by number of metapotential wells. The blue-colored region has a single well, the red-colored region has two wells, and the purple-colored region has three wells. The symbols $\times$ mark the line defined by the condition
    $\mathcal{U}(\mathcal{Q}_{\times\mathrm{min}}) =\mathcal{U}(\mathcal{Q}_{0}) = 0.$}. {The side plots represent the effective potential at the points indicated in the phase diagram. Here $K/(\kappa+\gamma)=-3.12\times10^{-5}$.} }
    \label{fig:phase_diagram}
\end{figure}

To examine in more detail the features of the phase diagram, we plot the effective potential in Fig. \ref{fig:phase_diagram}. We find the minima and maxima by examining the solutions of $\partial_{\mathcal{Q}}\mathcal{U}(\mathcal{Q})=0$. 

{In the region colored in blue - defined by the conditions ($\Delta  <0$ and $|\alpha| <\alpha_{c}(\Delta )$) or by ($\Delta >0$ and $|\alpha|<\alpha_{c}(0)$) -  the effective potential has a single global minimum at $\mathcal{Q}_{0}=0$.}

In the regions {colored in red or in purple,} where $|\alpha| > \alpha_{c}(0)$ and $\Delta > 0$, or where $|\alpha| > \alpha_{c}(\Delta )$, for $\Delta <0$ we get two (additional) minima at 
\begin{equation}
\mathcal{Q}_{\rm min} = \pm \frac{1}{\sqrt{6|K|}}\sqrt{\Delta + \sqrt{|\alpha|^2 - \alpha_{c}(0)^2}}. \label{eq:twomin}
\end{equation}
When inserted in the expression of the reduced effective potential Eq. (\ref{eq:effred}) we find
\begin{equation}
	\mathcal{V}(\mathcal{Q}_{\rm min}) = |K| \mathcal{Q}^4_{\rm min} \left( \frac{\Delta}{2} - \sqrt{|\alpha|^2 - \alpha_{c}(0)^2}\right)	
   \label{eq:valqunim}.
\end{equation}

Finally, in the {purple-colored} region where $\Delta > 0$ and $\alpha_{c}(\Delta )>|\alpha| > \alpha_{c}(0)$, we obtain two more local maxima at
\begin{equation}
\mathcal{Q}_{\rm max} = \pm \frac{1}{\sqrt{6|K|}}\sqrt{\Delta - \sqrt{|\alpha|^2 - \alpha_{c}(0)^2}}. \label{eq:twomax}
\end{equation}
in addition to the two minima at $\mathcal{Q}_{\rm min}$ from Eq. (\ref{eq:twomin}).
This results in 
\begin{equation}
\mathcal{V}(\mathcal{Q}_{\rm max}) = |K| \mathcal{Q}^4_{\rm max} \left( \frac{\Delta}{2} + \sqrt{|\alpha|^2 - \alpha_{c}(0)^2}\right)	,
\end{equation}

Furthermore, we can immediately calculate the curvatures of the effective potential around the extreme points relevant for the central well, $\mathcal{Q}= \mathcal{Q}_{0}=0$ and $\mathcal{Q}= \mathcal{Q}_{\rm max}$, the first being positive
\begin{equation}
\mathcal{V}''(\mathcal{Q}_{0}) = \frac{1}{2}  \left[\alpha_{c}(\Delta)^2 -|\alpha|^2\right]  >0 , \label{eq:deriv}
\end{equation}
and the other one negative
\begin{equation}
	\mathcal{V}''(\mathcal{Q}_{\rm max}) = - 12 |K|\mathcal{Q}_{\rm max}^2\sqrt{|\alpha|^2 - \alpha_{c}(0)^2} <0 .
\end{equation}
These curvatures are independent on the Kerr nonlinearity $K$ and they become zero at the phase boundary.

{A further approximation in this region can be obtained in the case when the operational point is not too far from the boundary of the phase transition, 
$\Delta \gtrsim \sqrt{|\alpha |^2 -\alpha_{c}(0)^2}$. As we will see, this is useful in answering the question: which terms in the effective potential are responsible for producing which parts of the three-well structure that we have obtained in the exact calculation?
For $\mathcal{Q}_{\rm max}$, in this limit we get the same form as Eq. (\ref{eq:twomax}), while $\mathcal{V}(\mathcal{Q}_{\rm max})$ 
can be approximated as 
\begin{equation}
\mathcal{V}(\mathcal{Q}_{\rm max}) \approx \frac{\Delta}{24|K|}\left( \Delta - \sqrt{|\alpha|^2 - \alpha_{c}(0)^2} \right). \label{eq:maxap}
\end{equation}
For the two wells in this approximation we have from Eqs. (\ref{eq:twomin},\ref{eq:valqunim})
that 
\begin{equation}
\mathcal{Q}_{\rm min} \approx \pm\sqrt{\frac{\Delta}{3|K|}}; ~~ \mathcal{V}(\mathcal{Q}_{\rm min}) \approx - \frac{\Delta}{18 |K|}. \label{eq_minap}
\end{equation}
We can now answer the question above: indeed, these approximate results can be obtained directly from the effective potential 
Eq.~(\ref{eq:effred}) in the following way.  First, for calculating $\mathcal{Q}_{\rm max}$ we can neglect the sextic term, based on the fact that we expect $\mathcal{Q}_{\rm max}$ to be small (metastable well narrow and shallow) near the phase transition. By maximizing the remaining quadratic and quarctic terms, we can obtain immediately Eq.~(\ref{eq:maxap}). Second, in the case of the two wells it is the quarctic and sextic terms that are relevant; by neglecting the quadratic term and minimizing the potential we get 
$\mathcal{Q}_{\rm max}$ and  $\mathcal{V}(\mathcal{Q}_{\rm max})$ as given by Eqs.~ (\ref{eq_minap}).}

The line with $\times$ markers corresponds to the situation in which all three minima are equal to the zero value of the effective potential. This means $\mathcal{V}(\mathcal{Q}_{\times\mathrm{min}}) =\mathcal{V}(0) = 0$, which yields the following equation for the pump amplitude
\begin{equation}
|\alpha_{\times}(\Delta )|^2 = \alpha_{c}(0)^2 + \frac{\Delta^2}{4}. \label{eq:star}
\end{equation}
By inserting this into the general expression Eq. (\ref{eq:twomin}) for $\mathcal{Q}_{\mathrm{min}}$ we get 
\begin{equation}
\mathcal{Q}_{\times\mathrm{min}} = \pm \sqrt{\frac{\Delta}{|K|}}.
\end{equation}
This line separates the region where $\mathcal{Q}=0$ is a local minimum and $\mathcal{Q}_{\mathrm{min}}$ are global minima from the region where the reverse occurs ($\mathcal{Q}=0$ is the global minimum and $\mathcal{Q}_{\mathrm{min}}$ are local minima). Because the order parameter for the global minimum changes from $0$ to $\mathcal{Q}_{\mathrm{min}}$, the line described by Eq. (\ref{eq:star}) is often referred to as the first-order transition. However, for finite-time experiments with pulsed parameters where we bring the system in the metastable $\mathcal{Q}=0$ local mininum and observe its decay into the global minimum, it is more natural to ascribe the first-order transition to the line $\alpha_{c}(\Delta )$.

To illustrate the main feature of each region, we present a few numerical values in Table \ref{tab:phasepoints}. {Note that $\mathcal{Q}_\text{max} \gg 1$, which provides a further consistency check of our semiclassical approach.} 
\begin{table}[h]
	\centering
	\begin{tabular}{ccccc}
		\toprule[1pt]
		Points & $\frac{\alpha}{\kappa+\gamma}$ & $\frac{\Delta}{\kappa+\gamma}$ & $\mathcal{Q}_\text{max}$ & $\frac{\mathcal{U}(\mathcal{Q}_\text{max})}{\kappa+\gamma}$ \\
		\midrule
		1  & 0.445 & -0.408 & N/A & N/A  \\
		2  & 0.496 & 0.297 & N/A & N/A  \\
		3  & 0.623 & 0 & N/A & N/A  \\
		4  & 0.506 & 0.111 & 13.5 & 0.272  \\
		5  & 0.556 & 0.408 & 29.6 & 20.3  \\
		6 & 0.572 & 0.556 & 38.6 & 71.6  \\
		7 & 0.519 & 0.445 & 40.4 & 60.0 \\
        8 & 0.594 & -0.371 & N/A & N/A \\
		\bottomrule[1pt]
	\end{tabular}
	\caption{{\bf Parameter values for the selected points in the phase diagram of Fig.\,\ref{fig:phase_diagram}.}}
	\label{tab:phasepoints}
\end{table}

{\it Discussion: relation with the standard mean-field approach.}

{Often in the literature the default method of analyzing the equations of motion is mean-field theory. In this approach, $\expval{a}$ is treated as a complex number, and terms like $\expval{{a^\dagger}^ma^n}$ are approximated as $\expval{a^\dagger}^m\expval{a}^n$ \cite{PhysRevA.94.033841,PhysRevA.103.033711}. Here we show that the mean-field approach produces results that are perfectly consistent with the effective potential method. We start by recalling that $a = (\mathcal{Q} + i \mathcal{P})/\sqrt{2}$
and $a^{\dag} = (\mathcal{Q} - i \mathcal{P})/\sqrt{2}$ therefore $\mathcal{N}= a^\dag a  = (\mathcal{Q}^2 + \mathcal{P}^2)/2 - 1/2$ is the particle number operator. In the mean-field approach, $\mathcal{P}$, $\mathcal{Q}$, and $\mathcal{N}$ are treated as classical variables. From Eqs. (\ref{eq:probedotq},\,\ref{eq:probedotp}) with $b=0$ and ignoring the noise terms we find that the stationarity condition $\dot{\mathcal{P}}= \dot{\mathcal{Q}} =0$ results in a system of equations that has either solution $\mathcal{P}=\mathcal{Q} = 0$ or admits a non-zero solution if the determinant of the matrix
\begin{align}
\left[
\begin{matrix}
|\alpha |\sin (\theta - \theta_{\rm P}) - \alpha_{c}(0)  &  - |\alpha|\cos (\theta - \theta_{\rm P} ) + \tilde{\Delta} \\
- |\alpha|\cos (\theta - \theta_{\rm P} ) - \tilde{\Delta} & -|\alpha |\sin (\theta - \theta_{\rm P}) - \alpha_{c}(0) 
\end{matrix}
\right] \nonumber
\end{align}
is zero. Here  $\tilde{\Delta} = \Delta + 12 K(\mathcal{N}+1/2)$ is the detuning renormalized by the nonlinearity.
Therefore we find that the stationary solutions are
\begin{eqnarray}
	\mathcal{N}_{0}& = & 0, \\
	\mathcal{N}_{\pm}& =  &\frac{\Delta\pm\sqrt{|\alpha|^2-\alpha_{c}(0)^2}}{12|K|}- \frac{1}{2}.
\end{eqnarray}
But now we have seen that typically $\mathcal{Q} \gg \mathcal{P}$, therefore we have approximately $\mathcal{N}_{+} \approx \mathcal{Q}_{\rm min}^2/2$ and $\mathcal{N}_{-} \approx \mathcal{Q}_{\rm max}^2/2$. These relations are verified immediately from Eq. (\ref{eq:twomin}) and Eq. (\ref{eq:twomax}), and of course we also have $\mathcal{N}_{0} = \mathcal{Q}_{0}$ trivially.}

{\it Kerr-free resonators.}
{Let us discuss here the situation where $K=0$. This case is directly relevant not only for optical realizations where the nonlinearities are very small, but also in superconducting circuits where SNAIL (Superconducting Nonlinear Asymmetric Inductive eLement) - based architectures such the ones used in traveling-wave parametric amplifiers \cite{PhysRevApplied.18.024063} can achieve a significant reduction of $K$ \cite{PhysRevApplied.10.054020}.}
 In this case the effective potential $\mathcal{U}(\mathcal{Q})$ becomes a parabola or an inverted parabola,    
\begin{equation}
\mathcal{U}(\mathcal{Q})  =  \frac{1}{2}s\mathcal{Q}^2.
\end{equation}
Here by $s\equiv s (|\alpha|, \Delta )$ we denote the curvature of the effective potential, that is, the following combination of pump strength $|\alpha|$ and detuning $\Delta$ 
\begin{equation}
s\equiv s (|\alpha|, \Delta ) = \partial_{\mathcal{Q}}^2 \mathcal{U} (\mathcal{Q})= \frac{\alpha_{c}(\Delta)^2 -|\alpha|^2}{\alpha_{c}(0) + |\alpha |}.
\end{equation}
{This curvature is the same as calculated before in Eq. (\ref{eq:deriv}) at $\mathcal{Q}_{0}=0$, only that now it extends for all $\mathcal{Q}$.}
{The boundary of the phase transition remains the same as in the $K\neq 0$ case, given by the function $\alpha_{c}(\Delta )$, as this quantity does not depend on the nonlinearity. All points below this line have $s>0$ (including our operational point) with a minimum at $\mathcal{Q}_{0}=0$, while above the line $s<0$ produces an unstable maximum. At resonance $\Delta =0$ and we have $s=(\kappa +\gamma )/2 - |\alpha|$, which we can interpret as a renormalization of the dissipation by the pump.}

\subsection*{Switching dynamics}

\begin{figure}
    \includegraphics[width=0.8\linewidth]{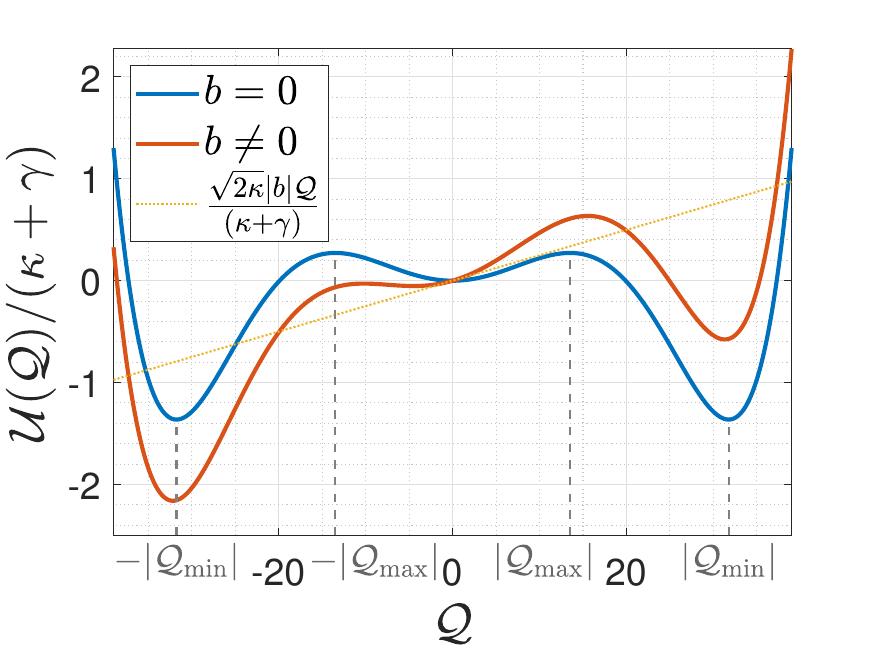}
    \caption{{\bf The effective potentials with probe on ($b\neq0$) and off ($b=0$).} {The potentials plotted here correspond to point \raisebox{.5pt}{\textcircled{\raisebox{-.9pt} {4}}} in Fig. ~1}: $\alpha/(\kappa+\gamma)=0.506$, $\Delta/(\kappa+\gamma)=0.111$ and $K/(\kappa+\gamma)=-3.12\times10^{-5}$. The non-zero term $|b|=\sqrt{2\times10^4} \sqrt{\text{Hz}}$ tilts the effective potential curve, making the barrier value $\mathcal{U}(\mathcal{Q}_{ \rm max}) - \mathcal{U}(0)$ smaller.
    At large $|b|$ values the barrier vanishes as $\mathcal{U}(-|\mathcal{Q}_{ \rm max}|) $ becomes negative.}
    \label{fig:metapotential_tilted}
\end{figure}

In this subsection, we present numerical and analytical results on the system's dynamics based on the Heisenberg-Langevin and Fokker-Planck equations. {We fix the phase of the pump $\theta_{\mathrm{P}}=0$ as reference, and we put $\theta = \pi/2$, which ensures that $\mathcal{Q}$ is slow and $\mathcal{P}$ is fast, as discussed {previously}.
Optimal detection happens for $\varphi =\theta /2$, when we have for the potential 
\begin{equation}
	\mathcal{U}_{b} (\mathcal {Q}) =  \mathcal{U} (\mathcal {Q}) + \sqrt{2\kappa}|b|\mathcal{Q} .\label{eq:U_tilted}
\end{equation}
{This potential is depicted in Fig.~\ref{fig:metapotential_tilted}.}
We set $\varphi =\pi/4$ throughout this subsection, with the exception of the last result where the phase-dependence is explicitly studied.}
{We start by presenting generic results corresponding to probe pulses applied at $t=0$ and measured at a time $t$, while in the end we apply these results to a realistic experimental sequence.}

\subsubsection*{Numerical check of separation of time-scales }

\begin{figure}
    \centering
    \includegraphics[width=.8\linewidth]{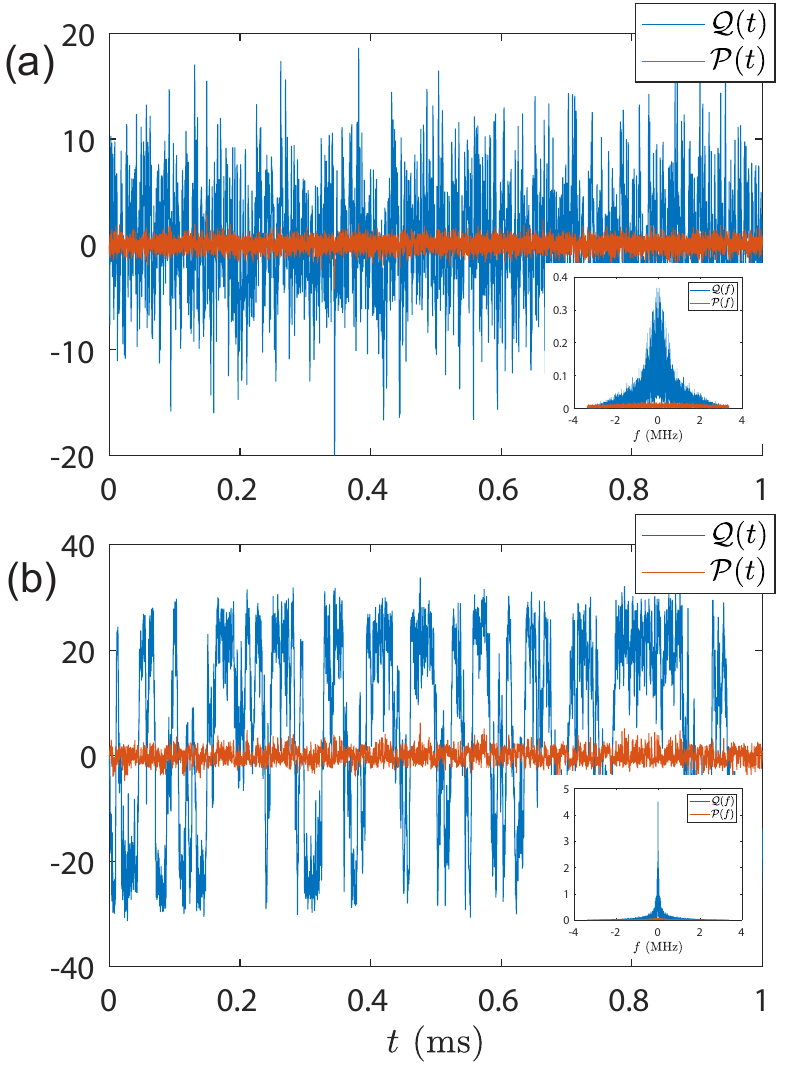}
    \caption{ {\bf Results for Heisenberg-Langevin equation simulation of the $\mathcal{Q},\mathcal{P}$ time-dependent quadratures.} \textbf{(a)} Below the {phase boundary} with  $|\alpha|/(\kappa+\gamma)=0.46, \Delta/2\pi=0.67$ MHz. \textbf{(b)} In the proximity of the {phase boundary} with $|\alpha|/(\kappa+\gamma)=0.505, \Delta/2\pi=0.67 $ MHz. {The insets show the amplitude spectra of the signals at a finite sampling rate such that it filters out the $\mathcal{P}$ quadrature while the slower $\mathcal{Q}$ remains.}}
    \label{fig:QP_046}
\end{figure}

{To start with, we demonstrate that the approximation of slow and fast dynamics for $\mathcal{Q}$ and $\mathcal{P}$ is self-consistent.} 
In Fig. \ref{fig:QP_046}, we show the numerical solution of the Heisenberg-Langevin equations (\ref{eq:probedotq}) and (\ref{eq:probedotp}) for the quadratures $\mathcal{Q}$ and $\mathcal{P}$ in the case of $b=0$. These results confirm our earlier assumptions: the slow quadrature $\mathcal{Q}$ is significantly larger than the fast quadrature $\mathcal{P}$.
{The insets show the corresponding amplitude spectra. To put in evidence the difference between time scales, we use a sampling rate of 6.738 MHz, which via the Nyquist–Shannon sampling theorem effectively filters the frequencies above, producing a low-pass cutoff at 3.369 MHz. We observe that the $\mathcal{P}$ quadrature is completely nullified due to fast-frequency dynamics, while $\mathcal{Q}$ has a non-zero spectrum due to frequency components lower than this cutoff.} 
This approximation becomes better as the parameters $\Delta, |\alpha|$ approach the critical threshold.  

\subsubsection*{Probability of switching}

{By solving either the Heisenberg-Langevin or the Fokker-Planck equations we can obtain the probability of finding the variable $\mathcal{Q}$ outside the interval $[-\mathcal{Q}_{\rm th}, \mathcal{Q}_{\rm th}]$ by integrating the probability density over the rest of the $\mathcal{Q}$-axis.
For the dark counts we write
\begin{align}
\label{eq:p_dark_FP}
p_{\texttt{dark}}(t) &\equiv p_{\texttt{dark}}((\mathcal{Q}<-|\mathcal{Q}_{\rm th}|) \cup (\mathcal{Q}>|\mathcal{Q}_{\rm th}|),  t) \nonumber \\
 & = 1-\int_{-\mathcal{Q}_{\rm th}}^{\mathcal{Q}_{\rm th}} W_{b=0}(\mathcal{Q},t) d\mathcal{Q},
 \end{align}
and for the case when an input field is present we have
\begin{align}
\label{eq:P_1+_FP}
  P_{\texttt{1\raisebox{-0.5ex}{+}}} (t) &\equiv P_{\texttt{1\raisebox{-0.5ex}{+}}} ((\mathcal{Q}<-|\mathcal{Q}_{\rm th}|) \cup (\mathcal{Q}>|\mathcal{Q}_{\rm th}|,
t) \\
& =1 -
	\int_{-\mathcal{Q}_{\rm th}}^{\mathcal{Q}_{\rm th}} W_{b}(\mathcal{Q},t) d\mathcal{Q}. 
\end{align}}

For characterizing the detector it is useful to define the concept of efficiency. 
Following the general quantum theory rules for non-number-resolving ideal detectors, for two states, denoted here by $\texttt{0}$ and $\texttt{1\raisebox{-0.5ex}{+}}$, we have the associated {positive operator-valued measure (POVM)} operators that are \cite{Kok2000,RevModPhys2007,migdallSinglephotonGenerationDetection2013}
\begin{eqnarray}
	\Pi_{\texttt{0}} &=& \sum_{n=0}^{\infty}(1-\eta )^{n}|n\rangle\langle n| ,\\
	\Pi_{\texttt{1\raisebox{-0.5ex}{+}}} &=& \mathbb{I} - \Pi_{0} ,
\end{eqnarray}
resulting in probabilities $p_{\texttt{0}}= {\rm Tr}\{\Pi_{\texttt{0}} \rho\} = \sum_{n=0}^{\infty}(1-\eta )^{n} P_{n}$ and $p_{\texttt{1\raisebox{-0.5ex}{+}}}= {\rm Tr}\{\Pi_{\texttt{1\raisebox{-0.5ex}{+}}} \rho\}= 1 - p_{\texttt{0}}$, where $\rho$ is the state at the input of the detector with photon number probabilities $P_{n} = {\rm Tr}\{ |n\rangle\langle n| \rho \} $
\cite{Kok2000,RevModPhys2007,migdallSinglephotonGenerationDetection2013}. 
On top of these, we should add the non-ideality presented by the dark counts, which are considered to be statistically independent, $1- P_{\texttt{1\raisebox{-0.5ex}{+}}} =(1-p_\texttt{dark})p_{\texttt{0}}$, 
or 
\begin{equation} \label{equ:p1+}
 p_{\rm 1_+}= \frac{P_{\texttt{1\raisebox{-0.5ex}{+}}} - p_\texttt{dark}}{1- p_\texttt{dark}}.
\end{equation}

Consider now a coherent state at the input with $\bar{n} = |v|^2 \tau$ photons in the pulse of duration $\tau$. Then 
$p_{0} = \exp (-\eta \bar{n}) = \exp (-\eta |b|^2 \tau)$, which yields
\begin{equation}
	\eta =\frac{1}{\bar{n}}\ln\frac{1-p_\texttt{dark}}{1-P_{\texttt{1\raisebox{-0.5ex}{+}}}},
	\label{eta}
\end{equation}
as the definition of the coherent-state efficiency as a function of the experimental or numerical results $p_\texttt{dark}$ and $P_{\texttt{1\raisebox{-0.5ex}{+}}}$. This definition is consistent since it can be shown numerically that the decay outside the central well can be approximated by decay rates $\Gamma_{\texttt{dark}}$ and $\Gamma_{b}$, therefore $p_{\texttt{dark}}(\tau)  = 1-\exp [-\Gamma_{\texttt{dark}}\tau ]$ and $P_{\texttt{1\raisebox{-0.5ex}{+}}} (\tau) =  1- \exp [-\Gamma_{b}\tau]$, which in turn define $\eta$ via $\Gamma_{b} -\Gamma_{\texttt{dark}} = \eta |b|^2$.

\subsubsection*{Analytical results for Kerr-free resonators}

{In the case $K=0$
the system can also function as a detector since the perturbation produced by the input field will change the probability of finding 
$\mathcal{Q}$ outside the interval $[-\mathcal{Q}_{\rm th}, \mathcal{Q}_{\rm th}]$.}

The Heisenberg-Langevin equation
$\dot{\mathcal{Q}} =	- \partial_{\mathcal{Q}}\mathcal{U}_{b}(\mathcal{Q}) - \sqrt{\kappa + \gamma} \xi_{\mathcal{Q}_{\rm in}}$ reads in this case:
\begin{equation}
\dot{\mathcal{Q}} =	- s \mathcal{Q}
-\sqrt{2\kappa} |b| - \sqrt{\kappa + \gamma} \xi_{\mathcal{Q}_{\rm in}},
\label{eq:HL}
\end{equation}
{describing a generalized Ornstein–Uhlenbeck process.} These equations can be solved in the following way. Let $\mu (t)= \langle \mathcal{Q}(t)\rangle$, and since the fluctuations average to zero we have
\begin{equation}
\dot{\mu} =	- s \mu
-\sqrt{2\kappa} |b|,
\end{equation}
with solution
\begin{equation}
\mu (t) = \mu (0) e^{-st} + \frac{\sqrt{2\kappa}|b|}{s} \left( e^{-st} -1 \right) .\label{eq:mu}
\end{equation}
Let us denote $\delta \mathcal{Q} = \mathcal{Q} - \mu$. By integrating Eq. (\ref{eq:HL}) we get 
\begin{equation}
\delta \mathcal{Q}(t) = e^{-st} \delta \mathcal{Q}(0) - \sqrt{\kappa + \gamma}e^{-st} \int_{0}^{t}e^{s t'}\xi_{\mathcal{Q}_{\rm in}}(t') {dt'}.
\end{equation}
{This expression allows us to calculate the second-order correlation function, by employing $\langle  \xi_{\mathcal{Q}_{\rm in}}(t)  \xi_{\mathcal{Q}_{\rm in}}(t')  \rangle =(\bar{n}_{T}+ 1/2) \delta (t-t')$.
For the autocorrelation function at two times $t_1$, $t_2$, where $t_{1}<t_{2}$ we obtain 
\begin{align}
\langle \delta\mathcal{Q}(t_1 ) \delta \mathcal{Q}(t_2 )\rangle = & \frac{D}{s}e^{-s (t_{2}-t_{1})} \left(
1- e^{-2st_{1}}\right) +  \nonumber \\
& + e^{-s(t_{1}+t_{2})} \langle \delta Q^2 (0) \rangle ,
\end{align}
from which we can also obtain the}
fluctuations $\sigma^2(t) \equiv  \langle \delta \mathcal{Q}^2(t)\rangle$
\begin{equation}
\sigma^2(t) = \sigma^2(0) e^{-2st} - \frac{D}{s}\left( e^{-2st}-1\right).\label{eq:sigma}
\end{equation}

The corresponding normal distribution describing the probability of finding the particle at position $\mathcal{Q}$ at time $t$ is
\begin{equation}
    W_{b}(Q,t) = \frac{1}{\sqrt{2\pi\sigma^2(t)}}e^{-\frac{[Q-\mu (t)]^2}{2\sigma^2(t)}}.
    \label{eq:Ga}
\end{equation}

Alternatively, we can obtain the same result by solving the Fokker-Planck equation.
We can search for a solution of the Fokker-Planck equation $\partial_{t} W_{b} = \partial_{\mathcal{Q}} \left( W_{b} \partial_{\mathcal{Q}}  \mathcal{U}_{b} \right) + D \partial_{\mathcal{Q}}^{2}W_{b}$,
	 in the form of a Gaussian function as in Eq.~(\ref{eq:Ga}), namely  $W_{b}(Q,t) = (1/{\sqrt{2\pi\sigma^2(t)}})\exp[{-\frac{[Q-\mu (t)]^2}{2\sigma^2(t)}}]$
 with variance $v(t)=\sigma^2(t)$ and mean $\mu(t)$. By inserting this ansatz into the Fokker-Planck equation we find that these parameters should satisfy
\begin{align}
\dot{\mu} &= - s\mu - \sqrt{2\kappa}b ,\\
\dot{v} & = -2 s v + 2 D .
\end{align}
By solving these equations we obtain the time dependence of the mean is
\begin{equation}
\mu (t) = \mu (0) e^{-st} + \frac{\sqrt{2\kappa}|b|}{s} \left( e^{-st} -1 \right),
\end{equation}
and the variance is 
\begin{align}
\sigma^2(t) = & \sigma^2(0) e^{-2 st} - \frac{D}{s}\left(e^{-2 st} -1
\right) ,
\end{align}
which are the same results as Eqs. (\ref{eq:mu}, \ref{eq:sigma}) obtained from the Heisenberg-Langevin equation. 

{We also note that for $s>0$ and in the asymptotic limit $t\gg 1/(2s)$ we obtain $\mu(t) \approx -\sqrt{2\kappa}|b|/s$ and $\sigma^2(t)\approx D/s$, therefore the probabilities become constant in time.}

{
Finally, the probabilities $P_{\texttt{1\raisebox{-0.5ex}{+}}}$ and $p_\texttt{dark}$ evaluated at some time $t$ can be expressed in terms of the error function $\mathrm{erf}(z) = (2/\sqrt{\pi})\int_{0}^{z}e^{-x^2}dx$ as
\begin{align}
P_{\texttt{1\raisebox{-0.5ex}{+}}} &= 1 - \frac{1}{2}\mathrm{erf}\left(\frac{\mathcal{Q}_{\mathrm{th}} + \mu}{\sqrt{2}\sigma}\right) - \frac{1}{2}\mathrm{erf}\left(\frac{\mathcal{Q}_{\mathrm{th}} - \mu}{\sqrt{2}\sigma}\right), \\
p_\texttt{dark} &= 1 - \mathrm{erf}\left(\frac{\mathcal{Q}_{\mathrm{th}}}{\sqrt{2}\sigma}\right) .
\end{align}
}

If we operate close to the threshold, $|\alpha |\approx \alpha_{c}(\Delta )$, then $s\approx 0$ and $\mathcal{U}(\mathcal{Q}) \approx 0$ and 
\begin{align}
\sigma^2(t) \approx \sigma^2(0) + 2Dt ,\\
\mu(t) \approx \mu (0) -\sqrt{2\kappa}|b|t .
\end{align}
In this situation, and with initial conditions $\sigma (0) = 0$, $\mu (0) =0$, we recover
from Eq. (\ref{eq:Ga}) the standard Fick's law for a Brownian particle under the action of a constant force, {in the form 
\begin{equation}
    W_{b}(Q,t) = \frac{1}{\sqrt{4\pi Dt}}e^{-\frac{(Q+\sqrt{2\kappa}|b|t )^2}{4Dt}}.
\end{equation}}
{For these processes one can find analytically the probability density function of the first times at which the particle reaches a value $\mathcal{Q}$.
This is known as first-passage time density probability,
and in the present context its variance can be interpreted as a measure of the detector jitter. Since with our convention the effective potential is tilted to the left, consider a point $\mathcal{Q}<0$: the times $t$ at which the particle first reaches this point are distributed according to the inverse Gaussian function,
\begin{equation}
\mathrm{IG} \left[ \frac{|\mathcal{Q}|}{\sqrt{2\kappa }|b|},\frac{\mathcal{Q}^2}{2D} \right] = \frac{|\mathcal{Q}|}{\sqrt{4\pi D t^{3}}}e^{-\frac{(Q+\sqrt{2 \kappa }|b|t)^2}{4Dt}},
\end{equation}
with mean $\mathbb{E}[t] = |\mathcal{Q}|/\sqrt{2\kappa}|b|$ and variance 
$\mathrm{Var}[t] = 2 |\mathcal{Q}|D/(\sqrt{2\kappa |}b|)^3$.}

\subsubsection*{{Numerical results}}

In our semiclassical method, we have represented the quantum input states as classical stochastic variables. The Heisenberg-Langevin equations Eqs. (\ref{eq:probedotq},\,\ref{eq:probedotp}) define a standard Wiener process with a Wiener increment $\sim \sqrt{dt} \mathbf{N}[0,1]$, where $\mathbf{N}[0,1]$ is a Gaussian random variable. {These equations can be solved by standard numerical techniques (see Methods).}

{For the timing of the pulses we follow the protocol described in detail in Ref. \cite{petrovnin_microwave_2024}: we start with the system in equilibrium with the pump switched off, then we ramp it up fast to bring the device to the operational point
\raisebox{.5pt}{\textcircled{\raisebox{-.9pt} {4}}} from Fig. \ref{fig:phase_diagram}. To avoid fast transients, we wait a short time ($0.23$ $\mu$s hereafter) before coupling in the probe field to be detected (with a duration of 1 $\mu$s in the following). In the presence of the probe field the effective potential becomes tilted, see Fig.~\ref{fig:metapotential_tilted}. We record the quadratures at each time $t$, and decide that the system has switched or not into one the outer effective potential wells depending on whether $|\mathcal{Q}|$ is larger or smaller than a suitably chosen threshold value $\mathcal{Q}_{\rm th}$. Eventually we turn off the pump, allowing the system to equilibrate and we repeat the process multiple times. This allows us to determine the probabilities of switching.}

\begin{figure}
    \centering
    \includegraphics[width=1\linewidth]{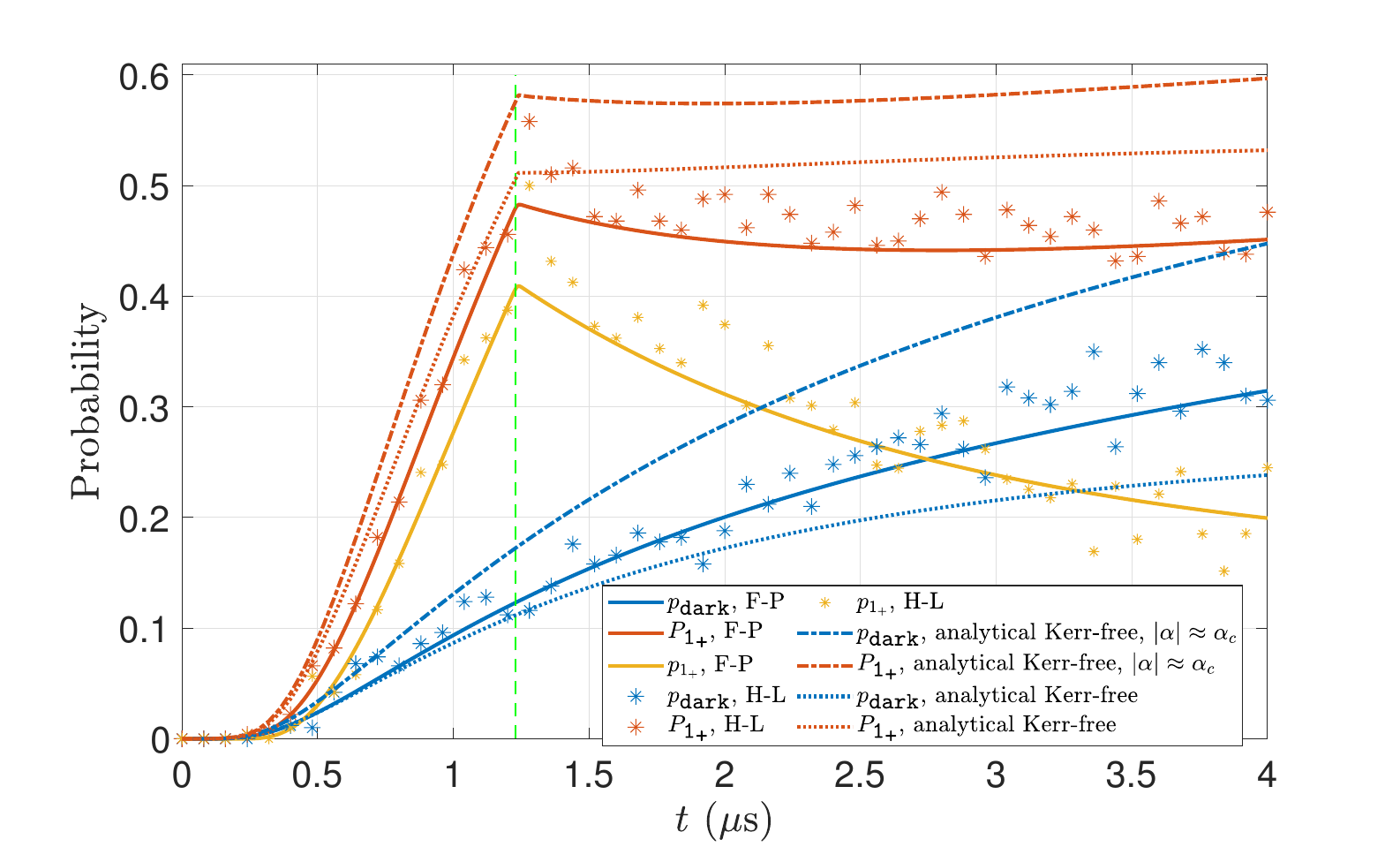}
    \caption{{\bf Numerical simulations and analytical results for the Heisenberg-Langevin and Fokker-Planck equations.} {The numerical results} for the Heisenberg-Langevin (star markers{, H-L in the legend}) and Fokker-Planck (solid lines{, F-P in the legend}) equations as a function of the measurement time $t$ in the range [0,4]$\,\mu$s {are obtained} {for a coherent probe pulse} with {duration of 1} $\mu$s (from 0.23 $\mu$s to 1.23 $\mu$s) and amplitude $|b|=10^3 \sqrt{\text{Hz}}$. {For the Heisenberg-Langevin equations we perform 250 runs for each point.} The vertical green dashed line denotes the falling edge of the probe pulse signal at $1.23\,\mu$s. The quadrature threshold is taken as $\mathcal{Q}_{\rm th}=7$. The other parameters correspond to point \raisebox{.5pt}{\textcircled{\raisebox{-.9pt} {4}}} from Fig. \ref{fig:phase_diagram}: $\alpha/(\kappa+\gamma)=0.506$, $\Delta/(\kappa+\gamma)=0.111 $ (or $\Delta/(2\pi)=0.748$\,MHz {with $\kappa/2\pi=4.44$\,MHz and $\gamma/2\pi=2.30$\,MHz}), and the Kerr coefficient is $K/(\kappa+\gamma)=-3.12\times10^{-5}$. The analytical results in the Kerr-free regime  at this operational point are presented with dotted lines. In addition, for the Kerr-free regime we show with dash-dotted lines the analytical results given for $|\alpha| \approx \alpha_c (\Delta )$, where $\Delta/(2\pi)=0.748$\,MHz, {in other words for a point on the phase transition boundary at the same detuning as \raisebox{.5pt}{\textcircled{\raisebox{-.9pt} {4}}}.} }
    \label{fig:FP_p_results2}
\end{figure}

{In Fig.~\ref{fig:FP_p_results2} we put together the numerical and analytical results for the case of coherent state input. The results obtained from the Heisenberg-Langevin Eqs. (\ref{eq:probedotq},\,\ref{eq:probedotp}) are compared to the solution of the Fokker-Planck Eq. (\ref{eq:FPb}).
For a finite Kerr nonlinearity, \raisebox{.5pt} we obtain the relevant 
probabilities $p_\texttt{dark}$, 
$P_{\texttt{1\raisebox{-0.5ex}{+}}}$, $p_{\texttt{1\raisebox{-0.5ex}{+}}}$ as a function of time, calculated for the sequence described above at the operational point
{\textcircled{\raisebox{-.9pt} {4}}}.
We can see that the effective potential theory that leads to the Fokker-Planck equation Eq.~(\ref{eq:FPb}) produces results that are close to those obtained from the Heisenberg-Langevin equations Eqs. (\ref{eq:probedotq},\,\ref{eq:probedotp}). This confirms that the approximations used in deriving the Fokker-Planck equation are satisfied.}

{The case $K=0$ for the same operational point \raisebox{.5pt}{\textcircled{\raisebox{-.9pt} {4}}} is shown with dotted lines. For this point ($\alpha=1.012\alpha_c$ and $\Delta /2\pi  = 0.748$ MHz) resulting in $s/2\pi=42.5$ kHz and with $D/2\pi = 1.685$ MHz therefore $D/s = 39.6$. Since $s>0$, this implies that for $t\gg1/(2s)=1.86$ $\mu$s the variance becomes asymptotically $\sigma^2(t)=D/s$ and the dark state probability at long times becomes $p_{\rm dark}=0.27$. 
We observe that 
$P_{\texttt{1\raisebox{-0.5ex}{+}}}$ increases while $p_\texttt{dark}$ decreases, which would result in a better detector. To understand why is it so, recall that the curvature of the effective potential does not depend on $K$ at $\mathcal{Q}_{0}=0$, thus the $K\neq 0$ and $K=0$ cases will have the same dynamics near this point. But, as the system diffuses and departs from $\mathcal{Q}_0$, the negative $~\mathcal{Q}^4$ contribution in the $K\neq 0$ case starts to kick in, lowering the potential and increasing $p_\texttt{dark}$. Finally, to achieve the condition $\alpha = \alpha_{c}(\Delta)$ or $s=0$ we increase $|\alpha|$ at the same detuning $\Delta$ until we reach the boundary of the phase transition.  The results are shown with dash-dotted lines. While $P_{\texttt{1\raisebox{-0.5ex}{+}}}$ increases, we also see that the dark count probability is higher compared to the previous cases, which means that this situation does not bring a clear advantage.}

\begin{figure}
    \centering
    \includegraphics[width=1\linewidth]{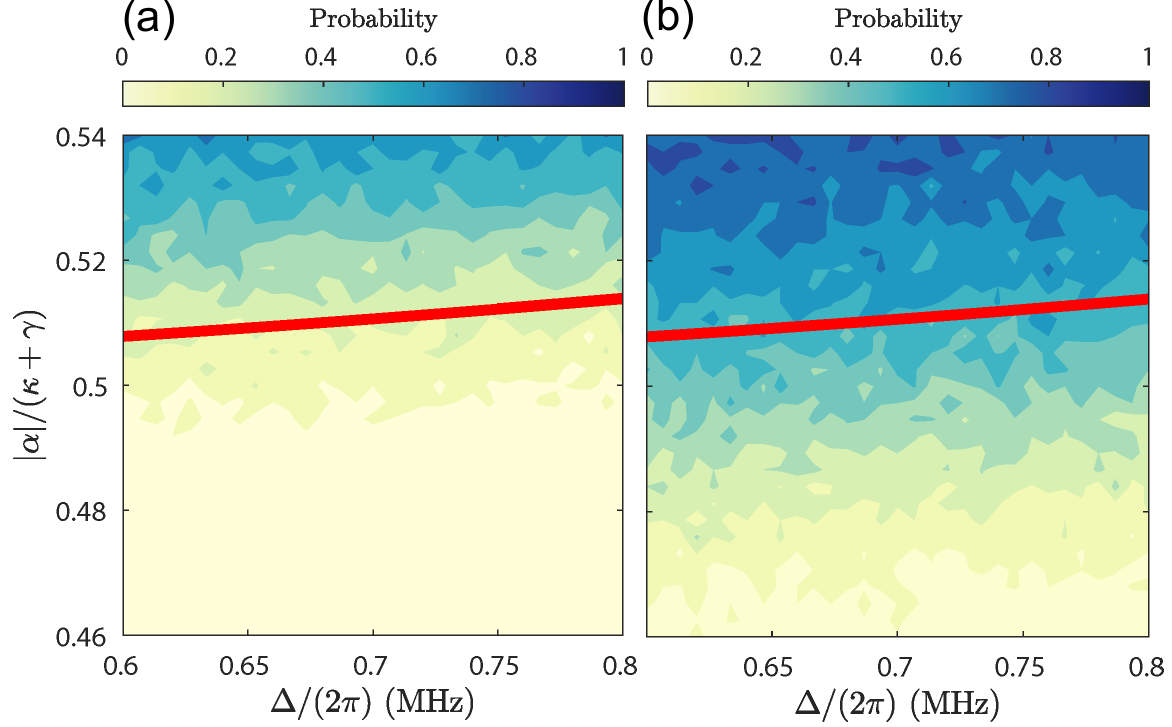}
    \caption{{\bf Experimentally-measurable switching probabilities for various 
    ($|\alpha|,\Delta$) parameters.} 
    The switching probabilities \textbf{(a)} $p_\texttt{dark}$ and $\textbf{(b)}$ $   P_{\texttt{1\raisebox{-0.5ex}{+}}}$ are calculated from the Heisenberg-Langevin equations with parameters in the range $|\alpha|/(\kappa + \gamma) \in [0.46,0.54], \Delta/ 2\pi \in [0.6\,\text{MHz},0.8\,\text{MHz}]$. {We perform 100 simulation runs for each ($|\alpha|,\Delta$) point.} The red lines marks the critical threshold $\alpha_{\rm c}(\Delta)$. The other parameters are $K/(2\pi)=-0.22$ kHz, $\gamma/(2\pi)=2.3\,\text{MHz}$ and $\kappa/(2\pi)=4.44\,\text{ MHz}$, and the signal is pulsed from $0.23$ to $1.23\,\mu\text{s}$ with $|b|=10^{3}\sqrt{\text{Hz}}$. }
    \label{fig:Sprobs-comparison}
\end{figure}

By numerically solving Eqs. (\ref{eq:probedotq},\,\ref{eq:probedotp}) for various parameters $\alpha$ and $\Delta$ we obtain the switching probabilities $p_\texttt{dark},
P_{\texttt{1\raisebox{-0.5ex}{+}}}$ at various points on the phase diagram near the phase transition, as shown in Fig. \ref{fig:Sprobs-comparison}. {These probabilities are measured at $t=1.23~\mu$s.} The coherent probe signal increases the switching probability in the selected parameter subspace, with larger impact from the coherent probe. The parameter subspace considered is near the border of the first-order phase transition with $\Delta > 0$, which is highly sensitive to small photon numbers. This claim is further supported by numerical results, which are obtained with $\Bar{n}=1$ for non-vacuum probe signal.

\begin{figure}
    \centering  \includegraphics[width=1\linewidth]{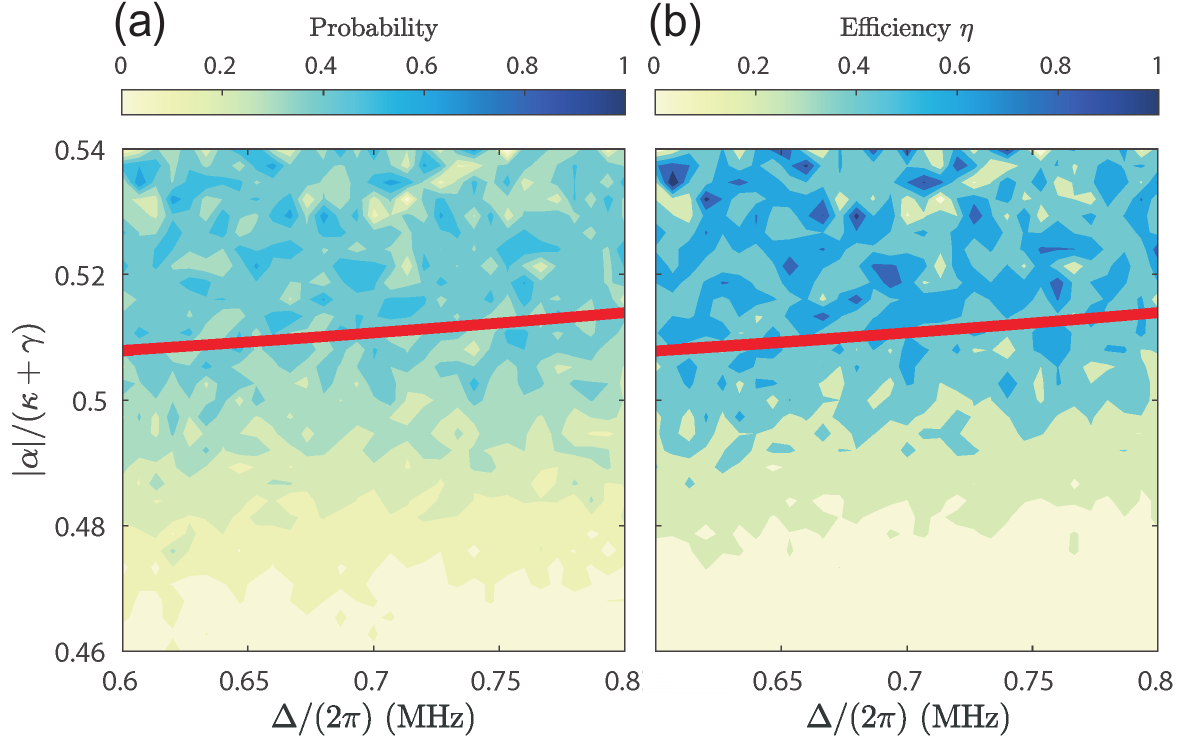}
    \caption{ {\bf The efficiency and the detection probability in the absence of dark counts.}
    {\bf (a)} The photon detection probability $p_{\rm 1_+}$ obtained from  the switching probability presented in Fig.~\ref{fig:Sprobs-comparison} by using Eq. (\ref{equ:p1+}). \textbf{(b)} The corresponding efficiency $\eta$ from Eq. (\ref{eta}) with $\bar{n}=1$. The red lines show the critical threshold $\alpha_{\rm c}(\Delta)$. } \label{efficiencies_plot_num}
\end{figure}

{Next, from Eq. (\ref{equ:p1+}) and with the data from Fig.~\ref{fig:Sprobs-comparison} we calculate the probability $p_{\texttt{1\raisebox{-0.5ex}{+}}}$, corresponding to a finite-efficiency detector without dark counts. The detection efficiency $\eta$ can be extracted by using Eqs. (\ref{eta}) with $\bar{n} = |b|^2 \tau  = 1$. The numerically obtained  $p_{\texttt{1\raisebox{-0.5ex}{+}}}$ and $\eta$ are shown in Fig. \ref{efficiencies_plot_num}. }
{Note that the figures in Fig.~\ref{fig:Sprobs-comparison} and Fig. ~\ref{efficiencies_plot_num} depend in general on the chosen readout time and the threshold value. This freedom can be leveraged when considering a specific application. For the values used here, $\mathcal{Q}_{\rm th} = 7$ and for the detection at the end of the pulse at $t = 1.23 \mu$\,s, we obtain at the operational point $p_\texttt{dark} \approx 0.14$, $P_{\texttt{1\raisebox{-0.5ex}{+}}} \approx 0.52$, and $\eta \approx 0.58$. Alternatively, by fitting the Fokker-Planck results in Fig. ~\ref{fig:FP_p_results2} we obtain the rates $\Gamma_{b} \approx 0.75$ MHz, $\Gamma_{\texttt{dark}} \approx 0.15$ MHz.,
yielding $\eta=(\Gamma_{b} -\Gamma_{\texttt{dark}}) / |b|^2 \approx 0.6$.}

\begin{figure}
    \centering  \includegraphics[width=0.9\linewidth]{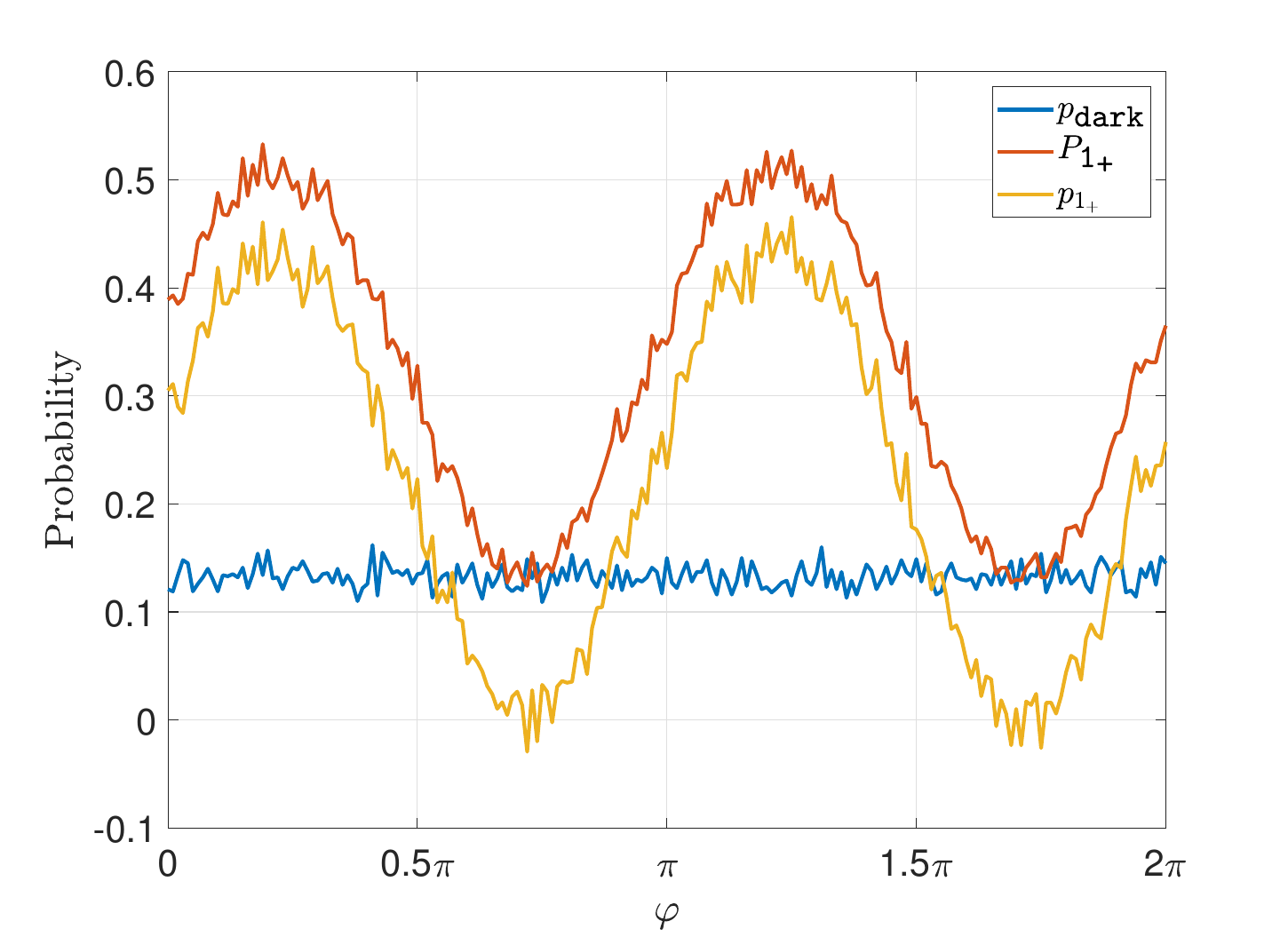}
    \caption{{\bf Phase dependence}. {{Numerical simulations of the Heisenberg-Langevin equations as a function of the coherent probe phase $\varphi$ in the range [0,2$\pi$].  The pulse has duration 1 $\mu$s (from 0.23 $\mu$s to 1.23 $\mu$s) and amplitude $|b|=10^3 \sqrt{\text{Hz}}$. The measurement is done at the falling edge of the probe pulse signal at $1.23\,\mu$s with the  quadrature threshold set to $\mathcal{Q}_{\rm th}=7$. 
    The other parameters are the same as in Fig. \ref{fig:FP_p_results2}.}}}
    \label{fig:phase_dependence}
\end{figure}

{Finally, we can demonstrate the role played by the phase $\varphi$ of the coherent probe. For the pulse and associated parameters studied in Fig.~\ref{fig:FP_p_results2} we plot the $\varphi$-dependent probabilities $P_{\texttt{1\raisebox{-0.5ex}{+}}}$, and $p_{\texttt{1\raisebox{-0.5ex}{+}}}$
in Fig.~\ref{fig:phase_dependence}, evaluated at the end of the pulse. As before, $\theta_{\mathrm{P}} = 0$ for reference and $\theta =\pi/2$.
Note that the results have periodicity $\pi$ due to the fact that we consider switching events in either one of the left or right well. The optimal case is $\varphi = \pi/4~ \mathrm{(mod~\pi)}$, which was the case considered throughout this paper. For $\varphi = \pi/2 ~\mathrm{(mod ~\pi)}$, the $|b|$-dependent term vanishes from Eq.~(\ref{eq:probedotq}) but not from Eq. ~(\ref{eq:probedotp}). However, we have argued {previously} that the latter is negligible. Indeed, from the simulations of the full equations we find that we reach the value $p_{\texttt{dark}}$ obtained in Fig.~\ref{fig:FP_p_results2} with the probe field completely off. This confirms the consistency of our approximations. 
}

\section*{Conclusions}
Using numerical and analytical methods, we have demonstrated that the parametric Kerr resonator operating near its critical threshold has the capability to detect very weak microwave signals.
The switching probabilities that constitute the output of such a detector can be obtained by solving the diffusive dynamics, either in the form of the Heisenberg-Langevin equations or by employing the Fokker-Planck equation in an effective potential. We have used stochastic numerical methods to solve the Heisenberg-Langevin {equation} and the results agree with those obtained from the Fokker-Planck equation. We also found analytical solutions for the Kerr-free case.
Our simulations and analytical results use realistic parameters obtained in the fabrication of typical superconducting amplifiers.
We note that for realizing single-photon detectors with high efficiency it may be necessary to reduce the Kerr nonlinearity coefficient. This can be done by engineering the effective potential by using SNAIL (Superconducting Nonlinear Asymmetric Inductive eLement) structures to achieve a smaller Kerr coefficient.

\section*{Aknowledgments}
We thank Niklas Toivonen for contributions to numerical simulations and data visualization in the early stages of this project as well as for help with drafting the manuscript. We are grateful to Pertti Hakonen, Charles Roques-Carmes, and Mark Dykman for inspiring discussions. This project has received funding from the European Union under Horizon Europe 2021-2027 Framework Programme, project MiSS (Microwave Squeezing with Superconducting (meta)materials) grant agreement ID: 101135868. We are also grateful for support from Saab, under a research collaboration agreement with Aalto University. This work was performed as part of the Finnish Center of Excellence in Quantum Technology QTF project no. 352925.

{\section*{Author contribution}{The conceptual framework of this problem was developed by KP and GSP. JW worked on the structure of the phase diagram, KP wrote the codes for the stochastic numerical simulations, and GSP was responsible for the analytical results.  All the authors discussed the findings presented in this work and contributed to the writing of the manuscript. }}

\section*{Data availability}
{All relevant data have been presented within the article.} 

\section*{Code availability}
The codes that were used for processing the data are
available from authors upon reasonable request. 

\section*{Competing interests} The authors declare no competing interests.

\section*{Methods}
\subsection* {Input-output theory}

The Heisenberg-Langevin equations Eqs. (\ref{eq:HL1},\,\ref{eq:HL2}) can be obtained also by including the probe field into the input external operator. With this interpretation we write
\begin{align}
  \dot{\mathcal{Q}} &= \frac{i}{\hbar}[H_{\rm sys}^{\text{(RWA)}},\mathcal{Q}]  - \frac{\kappa + \gamma}{2} \mathcal{Q}
  - \sqrt{\kappa}\mathcal{Q}_{\rm in}^{\rm [ext]} - \sqrt{\gamma} \xi_{\mathcal{Q}_{\rm in}^{\rm [int]}}
  \label{eq:HLL1}\\
    \dot{\mathcal{P}} &= \frac{i}{\hbar}[H_{\rm sys}^{\text{(RWA)}},\mathcal{P}]  - \frac{\kappa + \gamma}{2} \mathcal{P}
  - \sqrt{\kappa}\mathcal{Q}_{\rm in}^{\rm [ext]} - \sqrt{\gamma} \xi_{\mathcal{Q}_{\rm in}^{\rm [int]}}
  \label{eq:HLL2}
\end{align}
where the external input quadratures 
\begin{align}
	\mathcal{Q}_{\rm in}^{\rm [ext]} &= \sqrt{2}|b| \cos \left (\frac{\theta}{2} -\varphi\right) + \xi_{\mathcal{Q}_{\rm in}^{\rm [ext]}} , \\
	\mathcal{P}_{\rm in}^{\rm [ext]} &= \sqrt{2}|b| \sin \left (\frac{\theta}{2} -\varphi\right) + \xi_{\mathcal{P}_{\rm in}^{\rm [ext]}} ,
\end{align} 
result from writing $a_{\rm in}^{\rm [ext]} = |b|e^{i\left(\frac{\theta}{2}-\varphi \right)}
	+\frac{1}{\sqrt{2}}\left(\xi_{\mathcal{Q}_{\rm in}^{\rm [ext]}} +  i \xi_{\mathcal{P}_{\rm in}^{\rm [ext]}}\right)$.
{In this way the external input field operator is decomposed into a coherent component and a part that takes care of the thermal fluctuations, see also Ref.~\cite{RevModPhys.82.1155} for a related discussion.}

The external and internal noise operators 
($\xi_{\mathcal{Q}_{\rm in}^{\rm [ext]}}$, $\xi_{\mathcal{P}_{\rm in}^{\rm [ext]}}$) and ($\xi_{\mathcal{Q}_{\rm in}^{\rm [int]}}$, $\xi_{\mathcal{Q}_{\rm in}^{\rm [int]}}$) are assumed to be uncorrelated with each other and at the same temperature $T$, therefore
\begin{eqnarray}
	\langle  \xi_{\mathcal{Q}_{\rm in}^{\rm [ext]}}(t)  \xi_{\mathcal{Q}_{\rm in}^{\rm [ext]}}(t')  \rangle &=& (\bar{n}_{T}+ 1/2) \delta (t-t'), \nonumber \\
	\langle  \xi_{\mathcal{P}_{\rm in}^{\rm [ext]}}(t)  \xi_{\mathcal{P}_{\rm in}^{\rm [ext]}}(t')  \rangle &=& 
	(\bar{n}_{T}+ 1/2) \delta (t-t') , \nonumber \\
	\langle  \left[\xi_{\mathcal{Q}_{\rm in}^{\rm [ext]}}(t),  \xi_{\mathcal{P}_{\rm in}^{\rm [ext]}}(t') \right] \rangle &=& i \delta (t - t') , \nonumber
\end{eqnarray}
and 
\begin{eqnarray}
	\langle  \xi_{\mathcal{Q}_{\rm in}^{\rm [int]}}(t)  \xi_{\mathcal{Q}_{\rm in}^{\rm [int]}}(t')  \rangle &=& (\bar{n}_{T}+ 1/2) \delta (t-t'), \nonumber \\
	\langle  \xi_{\mathcal{P}_{\rm in}^{\rm [int]]}}(t)  \xi_{\mathcal{P}_{\rm in}^{\rm [int]}}(t')  \rangle &=& 
	(\bar{n}_{T}+ 1/2) \delta (t-t') , \nonumber \\
	\langle  \left[\xi_{\mathcal{Q}_{\rm in}^{\rm [int]}}(t),  \xi_{\mathcal{P}_{\rm in}^{\rm [int]}}(t') \right] \rangle &=& i \delta (t - t') . \nonumber
\end{eqnarray}
From here we can construct the total input noises used in Eqs. (\ref{eq:HL1},\,\ref{eq:HL2}) 
\begin{eqnarray}
\label{eq:totalQ_in_noise}
	\xi_{\mathcal{Q}_{\rm in}} &=& \frac{1}{\sqrt{\kappa + \gamma}}\left(\sqrt{\kappa} \xi_{\mathcal{Q}_{\rm in}^{\rm [ext]}} +
	\sqrt{\gamma} \xi_{\mathcal{Q}_{\rm in}^{\rm [int]} }\right) ,\\
    \label{eq:totalP_in_noise}
	\xi_{\mathcal{P}_{\rm in}} &=& \frac{1}{\sqrt{\kappa + \gamma}}\left(\sqrt{\kappa} \xi_{\mathcal{P}_{\rm in}^{\rm [ext]}} +
	\sqrt{\gamma} \xi_{\mathcal{P}_{\rm in}^{\rm [int]}} \right) .
\end{eqnarray}
with correlations represented by
\begin{eqnarray}
	\langle  \xi_{\mathcal{Q}_{\rm in}}(t)  \xi_{\mathcal{Q}_{\rm in}}(t')  \rangle &=& (\bar{n}_{T}+ 1/2) \delta (t-t'), \nonumber \\
	\langle  \xi_{\mathcal{P}_{\rm in}}(t)  \xi_{\mathcal{P}_{\rm in}}(t')  \rangle &=& 
	(\bar{n}_{T}+ 1/2) \delta (t-t') , \nonumber \\
		\langle  \left[\xi_{\mathcal{Q}_{\rm in}}(t),  \xi_{\mathcal{P}_{\rm in}}(t') \right] \rangle &=& i \delta (t - t') . \nonumber
\end{eqnarray}

By applying the general analysis of the power in the input external mode \cite{RevModPhys.82.1155} we find immediately that the contribution due to the coherent probe field is $\hbar \omega |b|^2$. Thus, the energy transferred to the system during a pulse of duration $\tau$
is $\hbar \omega |b|^2 \tau =\hbar \omega \bar{n}$, resulting in  $\bar{n} =  |b|^2 \tau$. {
This is consistent with the interpretation of the state characterizing the pulse as a displaced thermal state, with $\bar{n}_{T}$ average number of particles in the thermal component and a displacement $\sqrt{\bar{n}}$ produced by the coherent field. For these states, the total average number of particles is obtained as $\bar{n} + \bar{n}_{T}$.
We therefore have $\bar{n} =  |b|^2 \tau$ as the contribution of the coherent part.}

\subsection*{Numerical simulation of the Heisenberg-Langevin equation}
We {provide} here a pedagogical presentation of standard methods of discretization and numerical simulations of Langevin equations adapted to our problem. These techniques are then applied for various input signals leading to different distributions of the fluctuating term.

{The  differential Heisenberg-Langevin Eqs. (\ref{eq:probedotq},\,\ref{eq:probedotp}) have the generic form 
\begin{equation}
\frac{d}{dt}X (t) = f(X(t)) + \sqrt{2D}\eta (t) ,\label{eq:la}
\end{equation}
where $X(t)$ is a random variable, $f$ is the drift function, $D$ is the diffusion constant, and $\eta$ is a white noise with correlations 
\begin{equation}
\langle \eta (t) \rangle = 0,
\langle \eta (t) \eta (t')\rangle = \delta (t-t').
\end{equation}
We introduce a Wiener process as
\begin{equation}
\frac{d\mathcal{W}(t)}{d t} = \eta (t) ,
\end{equation}
and write accordingly Eq. (\ref{eq:la}) as
\begin{equation}
d X(t) = f(X(t)) + \sqrt{2D}d\mathcal{W}(t) .
\end{equation}
Let us now study the Wiener increment 
$d\mathcal{W}(t) = \mathcal{W}(t+dt)-\mathcal{W}(t)$. We have
\begin{equation}
\langle d\mathcal{W}(t) \rangle = 0 ,
\end{equation}
\begin{align}
\langle d\mathcal{W}(t)^2 \rangle = \int_{t}^{t+dt}\int_{t}^{t+dt} \delta (t'-t'') dt' dt'' = dt  .
\end{align} 
By using the central limit theorem, it follows that $d\mathcal{W}$ are normally distributed with variance $dt$ \cite{Gillespie}, in other words $dW(t) = \mathbf{N}(0,dt) = N(t) \sqrt{dt}$, where $N(t) = \mathbf{N}(0,1)$, is a sample value of the random normal variable $\mathbf{N}(0, dt)$ with average zero and  variance $dt$. 
We can then write
\begin{equation}
X(d + dt) = X(t) + f(X(t)) + \sqrt{2D}N(t)\sqrt{dt} ,\label{eq:num}
\end{equation}
which is a very useful update formula that can be used in numerics. 
We also have 
\begin{equation}
\eta (t) = \frac{d\mathcal{W}}{dt} = \frac{N(t)}{\sqrt{dt}}
= \frac{\mathbf{N}(0,1)}{\sqrt{dt}} = \mathbf{N}\left(0,\frac{1}{dt}\right) ,
\end{equation}
where in the last equality we used the properties of the normal distribution with respect to multiplication, $\mathbf{N}(0,1) = \sqrt{dt}\mathbf{N}(0, 1/(dt))$.
}

Eq.~(\ref{eq:num}) can be solved explicitly using standard numerical methods implemented in Matlab. We use the \textsc{ode45} solver, which employs the Dormand-Prince method from the Runge-Kutta family of ODE solvers.

\bibliography{Library.bib}

\end{document}